\documentclass{IEEEtran}
\usepackage{cite}
\usepackage{amsmath,amssymb,amsfonts}
\usepackage{algorithmic}
\usepackage{graphicx}
\usepackage{textcomp}
\usepackage{lipsum,mathtools,cuted}
\usepackage{xcolor}

\usepackage{float}
\usepackage{epsfig}
\usepackage{epstopdf}
\usepackage{subfigure,color}
\usepackage{dcolumn}
\usepackage{bm}
\usepackage{lineno}
\usepackage{accents}
\usepackage{multirow}
\usepackage[usestackEOL]{stackengine}
\usepackage{tabularx}
\usepackage{makecell}
\usepackage{dcolumn}% Align table columns on decimal point
\usepackage{bm}% bold math
\usepackage{booktabs} 
\UseRawInputEncoding

\def\BibTeX{{\rm B\kern-.05em{\sc i\kern-.025em b}\kern-.08em
    T\kern-.1667em\lower.7ex\hbox{E}\kern-.125emX}}
    
\def\e{\begin{equation}}
\def\f{\end{equation}}
\def\_#1{{\bf #1}}

\def\.{\cdot}

\def\=#1{\overline{\overline #1}}

\def\@#1{_{\rm #1}}
    
\begin{document}
\title{Discrete Impedance Metasurfaces for  6G Wireless Communications in D-Band}

\author{S.~Kosulnikov, X.~Wang, and S.~A.~Tretyakov,~\IEEEmembership{Fellow, IEEE}

\thanks{(Corresponding author: X.~Wang)}
\thanks{S.~ Kosulnikov and S.~A.~Tretyakov are with the Department of Electronics and Nanoengineering, Aalto University, FI-00079 Aalto, Finland. X.~Wang is with Institute of Nanotechnology, Karlsruhe Institute of Technology, 76131 Karlsruhe, Germany.(e-mail: sergei.2.kosulnikov@aalto.fi; xuchen.wang@kit.edu; sergei.tretyakov@aalto.fi).}}

\maketitle

\begin{abstract}
Engineering and optimization of wireless propagation channels will be one of the key elements of future communication technologies. Metasurfaces may offer a wide spectrum of functionalities for passive and tunable reflecting devices, overcoming fundamental limits of commonly used conventional phase-gradient reflectarrays and metasurfaces. In this paper， we develop an efficient way for the design and implementation of metasurfaces with high-efficiency anomalous reflector functionalities. The developed numerical method provides accurate, fast, and simple metasurface designs, taking into account non-local near-field interactions between array elements. The design method is validated by manufacturing and experimental testing of highly efficient anomalous reflectors for the millimeter-wave band. 
\end{abstract}


\maketitle

\section{Introduction}

\begin{figure}
\centerline{\includegraphics[width=0.8\columnwidth]{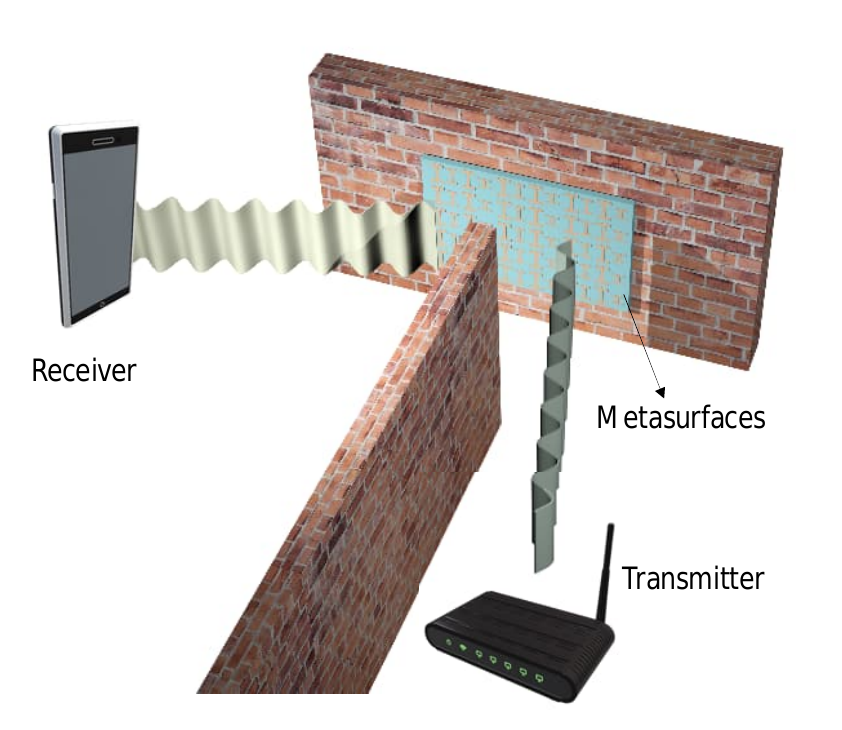}}
\caption{Application scenario of an anomalous reflector in 6G wireless communications.}
\label{fig:mainFigure}
\end{figure}

Novel wireless communication technologies are targeted for the need of growing data transfer rates, which leads to a shift to higher operational frequencies. The millimeter-wave (MM-wave) technologies become a basis for new generations of the wireless communication systems. 
However, communications at such high frequencies suffer from high free-space attenuation even in indoor scenarios. For this reason, it is necessary to use highly directive antennas, losing the advantages of multi-path propagation that usually provides connectivity even without a direct line of sight (LOS) between the antennas. 
One of the very noticeable techniques potentially solving this issue is based not on modifications of the source and/or receiver itself, but on modifications and optimizations of the propagation environment. Indeed, especially indoors, it is very attractive to realize a scenario where the signal will be efficiently relayed from some off-site objects. 
It is worth noting that repeater-like-devices do not have to be active:  even a passive but smart optimization of the signal propagation path can improve the communication channel by reflecting or transmitting wave beams towards the desired directions. We also note that this approach is in line with the ``green trends'' for efficient and ecological resource utilization. 
%This concept can be realized with the so-called reconfigurable intelligent surfaces (RIS) technology. Thus, efficient operating (including redirection, focusing and other more complex procedures) with the incident and reflected signals using RIS can bring a significant input to build qualitative and stable wireless communication channels. 
In this work, we consider a particular example of  passive anomalous reflectors that reflect incident waves to arbitrarily set reflection angles. 
%Such structure provides a scenario, when the operational signal reflection is different from the conventional mirror behavior. 
This functionality can be used to create an effective wireless link for  non-line-of-sight (NLOS) communications even for MM-wave systems based on the use of high-gain antennas.  
Figure~\ref{fig:mainFigure} illustrates the considered scenario, where the signal from the transmitter cannot reach the client receiver in the LOS mode due to a wall obstacle, but still a reliable wireless channel is set due to redirecting and focusing the reflected signal by an anomalously reflecting metasurface.

The research of anomalous reflectors has been actively going on from 60's. Perhaps the earliest anomalous reflectors are reflectarrays where each element separated by $\lambda/2$ distance from neighboring elements locally reflects at different phases, forming a desired reflection wavefront \cite{Berry_reflectarray, Huang2008reflectarray}.
However, for grazing  reflection angles, the device efficiency significantly drops. Later, this idea was implanted in arrays with subwavelength unit cells: phase-gradient metasurfaces \cite{Pozar_MS,yu2011light}. The reflection phase of each unit cell of the  metasurface (realized as a periodical lattice of cells) is linearly and periodically varying on the surface. 
Still, such phase-gradient metasurfaces inherit the main drawbacks of reflectarrays: high reflection efficiency is achievable only when the deviation from the usual reflection law is not large. At sharp reflection angles spurious scattering dramatically increases. 
Since 2016, it is recognized that low efficiency of a device operating far from the specular or retro-reflection regimes is caused mainly by the impedance mismatch of the incident and reflected waves \cite{estakhri2016wave,asadchy2016perfect}. 
For a given plane-wave incidence, considering only a single plane wave reflection, it is not enough to satisfy the local lossless boundary condition of the metasurface \cite{asadchy2016perfect}. In \cite{epstein2016synthesis}, it was pointed out that evanescent fields have to be engaged and optimized to realize a purely reactive boundary that perfectly directs an incident plane wave to an anomalous direction. 
Based on this general principle, different design methods have been proposed. For example, in \cite{kwon2018lossless,budhu2020perfectly}, the evanescent fields are optimized by ensuring local power conservation on the surface and finding a locally reactive surface. In contrast, the design methods in  \cite{diaz2017generalized, he2022perfect, asadchy2017eliminating} are based on the global power conservation where the structures of meta-atoms are collectively optimized within a super-cell. This method was recently used for the creation of anomalous reflectors in D-band~\cite{Kato2021}, where patch sizes in a super-cell were collectively optimized starting from the local periodic approximation as an initial guess, with good results.  
In parallel, the meta-grating method \cite{ra2017metagratings, rabinovich2018analytical} was suggested as another alternative to realize perfect anomalous reflection with a reduced number of meta-atoms.  

In this paper, we propose an effective design method for the realization of perfect anomalous reflection into arbitrary directions. As a basis for design, we select an impedance sheet placed on a grounded dielectric substrate. We first discretize the  impedance sheet into finite elements with uniform values of the sheet reactance. For this discretized structure we derive an analytical formula for all the  scattered Floquet modes of the metasurface. Then we use an inverse design optimization method to optimize the discrete impedances until the amplitudes (and possibly phases) of the scattered modes satisfy the design objectives. This method is different from the known impedance-based methods \cite{kwon2018lossless, wang2020theory}, where the sheet  impedance is first assumed to be continuous in the optimization process and  discretized into finite elements only after optimization. After discretization, such metasurfaces may not behave as perfect as expected from the theoretical results of the continuous-profile optimization. In contrast, in the proposed method, the sheet impedance is optimized already in the discretized form. Therefore, the performance does not suffer from degradations caused by surface discretization.
In addition, the proposed method is not limited by the requirement of the local power conservation on the surface: the input impedance seen at the input surface can take purely imaginary or complex values. The overall passivity of the device is automatically ensured due to  the passivity of the impedance sheet and the substrate material. 
Therefore, the optimized results encompass both realization possibilities (local or global power conservation), which is more general than in the earlier developed methods. 
In addition, using the same approach, it is possible to design other devices for wave shaping, such as multi-channel beam splitters with arbitrarily assigned power ratios.
% This method is based on Ref.~\cite{Wang2020}, where the surface impedance is continuous. The continuous surface impedance requires many subelements to achieve a reasonable discretization, whereas with the developed in this work method there is no need in such further discretization. 

This paper is organized as follows: in the second section we introduce the theoretical model of the developed numerical optimization method, in the third section we focus on a particular example implementing an anomalous reflector operating at D-band, the fourth section is focused on the experimental validation of the implemented anomalous reflector, the conclusions and discussion section is closing the paper.

\section{Discrete sheet impedance model for reflection optimization}

\begin{figure}
\centerline{\includegraphics[width=0.99\columnwidth]{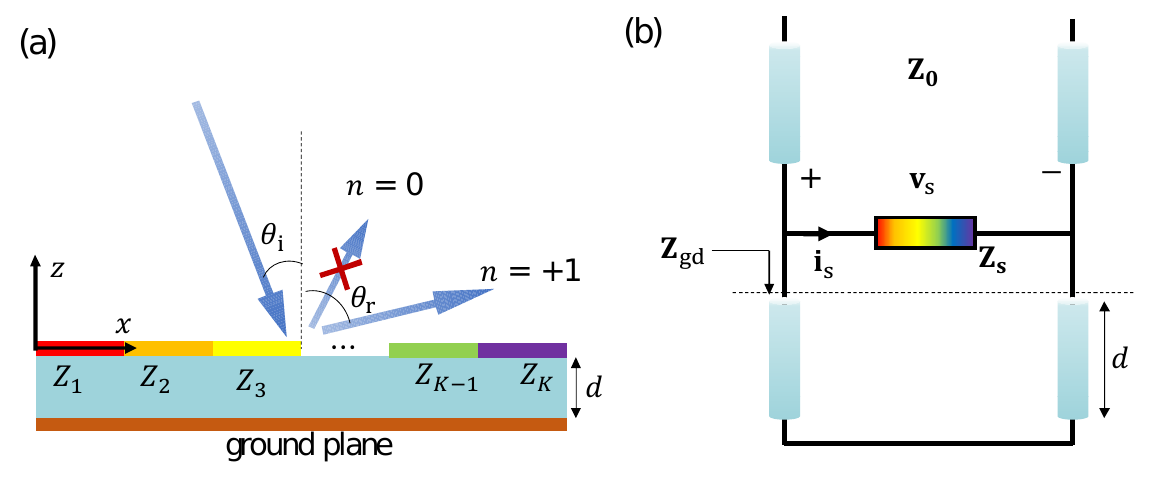}}
\caption{(a) Side view of a metasurface containing discrete impedance sheets on a grounded substrate. (b) Equivalent circuit of the metasurface.}
\label{fig:circuit and structure}
\end{figure}

The metasurface is composed of a discrete impedance array on top of a grounded dielectric substrate with permittivity $\epsilon_{\rm d}$ and thickness $d$. Figure~\ref{fig:circuit and structure}(a) shows a single unit cell of the metasurface with the spatial periodicity $D$ (the size of one super-cell).  
We discrete one super-cell into $K$  elements, as shown in Fig.~\ref{fig:circuit and structure}(a), where the central coordinate of the $m$-th cell is denoted as $x_m={(2m-1)D}/{2K}$.
The sheet impedance of the $m$-th subcell is denoted as $Z_m$ where $m\in[1, K]$. The discretized sheet impedance profile $Z_{\rm s}(x)$ is piece-wise homogeneous and can be viewed as a cascade of $K$ step functions.
As for any periodic function, it can be decomposed into Fourier series as 
\begin{equation}
    Z_{\rm s}(x)=\sum_{m=-\infty}^{+\infty}g_{m} e^{-jm\beta_{\rm M} x},  
    \label{Eq:Fourier_series_grid_impedance}
\end{equation}
where $g_m$ are the Fourier coefficients.
A plane wave illuminates the metasurface at the angle of $\theta=\theta_{\rm i}$. Due to the periodicity of the reflector, the scattered field is a sum of an infinite number of Floquet harmonics,  denoted by index $n$.
The tangential wavevector of the Floquet modes scattered from  periodical structures can be written as $k_{xn}=k_0\sin\theta_{\rm i}+n\beta_{\rm M}$, where $\beta_{\rm M}=2\pi/D$ is the spatial modulation frequency of the surface impedance, $n$ is the mode order,  $\theta_{\rm i}$ is the incident angle,  and $k_0=\omega_0\sqrt{\epsilon_0\mu_0}$ ($\omega_0$ is the incident-wave frequency) is the free-space wavenumber. 
Modes satisfying $|k_{xn}|<k_0$ propagate into the far-zone at the angles $\theta_n$, defined by \begin{equation}
    \sin\theta_n=\frac{k_0\sin\theta_{\rm i}+n\beta_{\rm M}}{k_0}.
    \label{Floquet_period}
\end{equation} 
For  $|k_{xn}|>k_0$, the modes are evanescent, exponentially decaying along the surface normal.

Next, based on the \textit{mode-matching} method, we aim to find the amplitudes of all scattered modes at the metasurface plane ($z=0$). For convenience of analysis, we use the transmission-line model shown in Fig.~\ref{fig:circuit and structure}(b). The sheet impedance is modeled as a shunt impedance in the transmission line. The grounded 
substrate is modeled as a shorted transmission line with its length equal to the substrate thickness. The current and voltage in the transmission-line model are analogous to the total tangential electric and magnetic fields on the surface ($z=0$). Unlike conventional transmission-line modeling of uniform metasurfaces, where the current and voltage only have one mode and can be treated as scalar numbers, in this case, the space-modulated metasurface excites infinitely many spatial modes (including propagating  and evanescent ones). Therefore, the current and voltage  in the transmission line are not  simple scalar numbers but infinite sums:
\begin{subequations}
    \begin{equation}
		I_{\rm s}(x)=\sum_{n=-\infty}^{+\infty}i_{\rm s}^n e^{-jk_{xn}x}  \label{Eq:current_space_modulation}
		\end{equation} 
and
	\begin{equation}
		\quad V_{\rm s}(x)=\sum_{n=-\infty}^{+\infty}v_{\rm s}^n e^{-jk_{xn}x}.\label{Eq:voltage_space_modulation}
	\end{equation}
\end{subequations}

The current and voltage on the impedance sheet must obey Ohm's law, $V_{\rm s}(x)=Z_{\rm s}(x)I_{\rm s}(x)$.
Substituting Eqs.~(\ref{Eq:current_space_modulation}), (\ref{Eq:voltage_space_modulation}), and (\ref{Eq:Fourier_series_grid_impedance}) into Ohm's law, we have
\begin{equation}
	\sum_{n=-\infty}^{+\infty}v_{\rm s}^n e^{-jk_{xn} x}=\sum_{n=-\infty}^{+\infty}\sum_{m=-\infty}^{+\infty}g_mi_{\rm s}^{n} e^{-jk_{x,n+m} x}. \label{Eq:admittance_expansion}
\end{equation}
By shifting the summation order $n$ to $n-m$ on  the right side of Eq.~(\ref{Eq:admittance_expansion}), we get an expression where both sides share the same basis $e^{-jk_{xn}x}$ which can be eliminated together with the summation over $n$. Therefore, Eq.~(\ref{Eq:admittance_expansion}) can be simplified as
\begin{equation}
	v_{\rm s}^n=\sum_{m=-\infty}^{+\infty}g_m i_{\rm s}^{n-m}.\label{Eq:current_voltage_admittance}
\end{equation}
One can see from Eq.~(\ref{Eq:current_voltage_admittance}) that every voltage harmonic is related to all current harmonics due to mode coupling.
Considering a finite number of Floquet modes from  $n=-N$ to $n=N$,  Eq.~(\ref{Eq:current_voltage_admittance}) can be written in a form of matrix multiplication:
\begin{equation}
	\begin{pmatrix}
	    v_{\rm s}^{-N}  \\
	    v_{\rm s}^{1-N} \\
	    \vdots   \\
	    v_{\rm s}^{+N}
	\end{pmatrix}=
	\begin{pmatrix}
	    g_{0} & g_{-1} & \cdots & g_{-2N} \\
	    g_{1} & g_{0} & \cdots & g_{1-2N} \\
	    \vdots  & \vdots  & \ddots & \vdots  \\
	    g_{2N} & g_{2N-1} & \cdots & g_{0}
	\end{pmatrix}
	\begin{pmatrix}
	    i_{\rm s}^{-N}  \\
        i_{\rm s}^{1-N} \\
	    \vdots   \\
	    i_{\rm s}^{+N}
    \end{pmatrix}.
    \label{Eq:impedance_matrix}
\end{equation}
Denoting the current and voltage arrays in Eq.~(\ref{Eq:impedance_matrix}) as $\mathbf{v}_{\rm s}$ and $\mathbf{i}_{\rm s}$, and the $(2N+1)$ dimensional impedance matrix  as $\mathbf{Z}_{\rm s}$, Eq.~(\ref{Eq:impedance_matrix}) can be written in a simple form:  $\mathbf{v}_{\rm s}=\mathbf{Z}_{\rm s} \cdot \mathbf{i}_{\rm s}$.

Similarly to the sheet impedance, the input impedance of the grounded substrate can also be written in matrix form. 
Since the substrate is spatially uniform and there is no mode coupling in the uniform substrate, the impedance of the grounded substrate  is a $(2N+1)$ dimensional \textit{diagonal} matrix $\mathbf{Z}_{\rm gs}$.  The $n$-th diagonal term is the input impedance of grounded substrate for the $n$-th scattering mode (note that we number the matrix columns and rows from $-N$ to $+N$ instead of from $1$ to $2N+1$):
\begin{equation}
   \mathbf{Z}_{\rm gs}(n,n)= \frac{1}{Z_{{\rm d},n}\tanh(jk_{zn}^{\rm d} d)}.\label{admittance matrix of transmission line}
\end{equation}
Here, $Z_{{\rm d}{,n}}$ is the wave impedance of the dielectric substrate material, and $k_{zn}^{\rm d}=\sqrt{\omega_0^2\epsilon_0\epsilon_{\rm d}\mu_0-k_{xn}^2}$ is the normal component of the wavevector in the dielectric substrate for the $n$-th mode. 
For TE polarized waves, $Z_{{\rm d},{n}}=(\mu_0\omega_0)/k_{zn}^{\rm d}$ and for TM polarized waves $Z_{{\rm d},{n}}=k_{zn}^{\rm d}/(\omega_0 \epsilon_0\epsilon_{\rm d})$. 
Details for derivation of (\ref{admittance matrix of transmission line}) can be found in Sec.~4 of the Supplementary Material of \cite{wang2020theory}.

The total input impedance of the metal-backed metasurface can be calculated as a  parallel connection of  the gradient penetrable impedance (characterized by $\mathbf{Z}_{\rm s}$) of the sheet and the input impedance of the metal-backed substrate $\mathbf{Z}_{\rm gs}$, i.e.,  $\mathbf{Z}_{\rm tot}=\mathbf{Z}_{\rm s}\rm || \mathbf{Z}_{\rm gs}$. 
After $\mathbf{Z}_{\rm tot}$ is determined, we can calculate all the scattered modes for a given incidence in terms of a reflection matrix. The reflection matrix is defined as
\begin{equation}
    \mathbf{\Gamma}_{\rm TM}=(\mathbf{Z}_{\rm tot}+\mathbf{Z}_{0})^{-1}\cdot  (\mathbf{Z}_0-\mathbf{Z}_{\rm tot}) \label{eq: gamma1}
\end{equation}
for TM-polarized incidence. Here,   $\mathbf{Z}_0$ is the wave impedance of free space, which is a diagonal matrix. The $n$-th element of $\mathbf{Z}_0$ has the same format with $Z_{{\rm d}, n}$, only assuming $\epsilon_{\rm d}=1$. The reflection matrix for TE-polarized incidence is 
\begin{equation}
    \mathbf{\Gamma}_{\rm TE}=(\mathbf{Y}_{\rm tot}+\mathbf{Y}_{0})^{-1} \cdot (\mathbf{Y}_0-\mathbf{Y}_{\rm tot}), \label{eq: gamma2}
\end{equation}
where $\mathbf{Y}_{\rm tot}=\mathbf{Z}_{\rm tot}^{-1}$ and $\mathbf{Y}_0=\mathbf{Z}_{\rm 0}^{-1}$ are the admittance matrices. The derivation details of Eqs.~(\ref{eq: gamma1}) and (\ref{eq: gamma2}) are presented in APPENDIX.
The reflection matrix relates the \textit{tangential } components of the incident and reflected fields, i.e., $\mathbf{E}_{\rm r}=\mathbf{\Gamma}_{\rm TE} \cdot\mathbf{E_i}$ for TE waves and $\mathbf{H}_{\rm r}=\mathbf{\Gamma}_{\rm TM} \cdot\mathbf{H_i}$ for TM waves. 
The incident and reflected tangential fields ($\mathbf{E}_{\rm i}$, $\mathbf{E}_{\rm r}$, $\mathbf{H}_{\rm i}$, and $\mathbf{H}_{\rm r}$) should be represented by $2N+1$-dimensional vertical vectors. The vector contains the complex amplitudes of the considered harmonics.

The above theory shows how to calculate all the scattered 
harmonics for a given set of discrete impedance sheets and
illumination waves. However, 
the design of metasurfaces is an inverse problem: for a given incidence and desired reflection harmonics, how to find a proper set of discrete impedance values? Since there is no analytical solution for this inverse problem, we use mathematical optimization. Next, we introduce the optimization principles. 
For a single TE-polarized plane wave incidence, $\mathbf{E}_{\rm i}$ can be written as
\begin{equation}
\mathbf{E}_{\rm i}=[0, \cdots, 0, 1, 0, \cdots, 0]^T,
\end{equation}
where the incident mode ($n=0$) is positioned in the middle of the array.
% Knowing the incident field $\mathbf{E}_{\rm i}$, the amplitudes of all scattered modes can be found by $\mathbf{E}_{\rm r}= \mathbf{\Gamma}\cdot \mathbf{E}_{\rm i}$. 
For a known illumination (a given $\mathbf{E}_{\rm i}$), our goal is to find a proper set of discrete grid impedances of an array that realizes the desired $\mathbf{E}_{\rm r}$. Here, we focus on the perfect anomalous reflection functionality with the incident power fully reflected to the $n=+1$ scattering order.  Therefore, the desired reflected field vector can be written as,
\begin{equation}
    \mathbf{E}_{\rm r}=[0, \cdots, 0, 0, A_{\rm obj}, \cdots, 0]^T,
\end{equation}
where $A_{\rm obj}=\sqrt{\frac{\cos\theta_{\rm i}}{\cos\theta_{\rm r}}}$ is the amplitude of reflection that ensures that all the incident power is directed to the anomalous direction \cite{asadchy2016perfect}. 
This goal is realized by optimizing the discrete impedance values $Z_1, Z_2, \cdots Z_K$ (purely reactive), or, equivalently to say, optimizing the reflection matrix $\mathbf{\Gamma}_{\rm TE}$, until the desired reflection mode amplitude is maximized  in the  reflection vector $\mathbf{E}_{\rm r}$. Here, we use the mathematical optimization tool available
in the MATLAB package to find the optimal values of the impedance sheets. 
In each trial of the optimization, MATLAB assumes an array of $Z_1, Z_2, \cdots, Z_K$ and calculates the reflected fields. Denoting the calculated amplitude of the desired reflection mode in each trial as $A_{\rm cal}$, the optimization goal is to find a proper set of $Z_1, Z_2, \cdots, Z_K$ that minimizes the  cost function defined as
\begin{equation}
    F(Z_1, Z_2, \cdots, Z_K)=|A_{\rm cal}-A_{\rm obj}|. \label{Eq:cost_function}
\end{equation}
Employing the MultiStart and $fmincon$ algorithms, MATLAB can search for the minimum value of $F$ in the multidimensional parameter space.

It is important to mention that this approach is quite general. It can be used for the design of Floquet metasurfaces with arbitrary power distributions among all the possible Floquet modes, e.g., beam splitters,  only by modifying the objective function in Eq.~(\ref{Eq:cost_function}). In addition, the desired reflection phases can also be set arbitrarily.

\section{Example of discretized impedance optimization}

\subsection{Design goals and optimization results}
We target to find designs of anomalous reflecting metasurfaces to realize the application scenario of 
Fig.~\ref {fig:mainFigure} at three potential operational frequencies for D-band communications: $f_0 = [144.75; 157.75; 170.90]$~GHz (the corresponding operational wavelengths $\lambda_0 = [2.0725; 1.9017; 1.7554]$~mm), marked as Designs 1, 2, and 3 in  Table~\ref{tab:MainTable}). The 
incident TE-polarized plane wave comes from the normal direction $\theta_{\rm i} = 0^{\circ}$, and the goal is to reflect it into the oblique direction $\theta_{\rm r} = 50^\circ$. 
Thus, the metasurface period is  $D = \lambda_0/\sin{\theta_{\rm r}}$. We discretize the impedance sheet into $K = 8$ elements, and assume a quartz substrate with the permittivity $\epsilon_{\rm d} = 4.2(1-j0.005) $ and thickness $d=209~\mu$m. In the optimization of the sheet reactances of the elements according to the cost function Eq.~\eqref{Eq:cost_function}, the allowed solutions of $Z_k$ were restricted to the range $[-2000j, +50j]$~Ohm. This constraint is introduced to make the actual implementations of the array elements easier, because large negative or positive reactances require some extreme geometries of metal elements that may be difficult or impossible to fabricate. 

\begin{figure}
\centerline{\includegraphics[width=0.8\columnwidth]{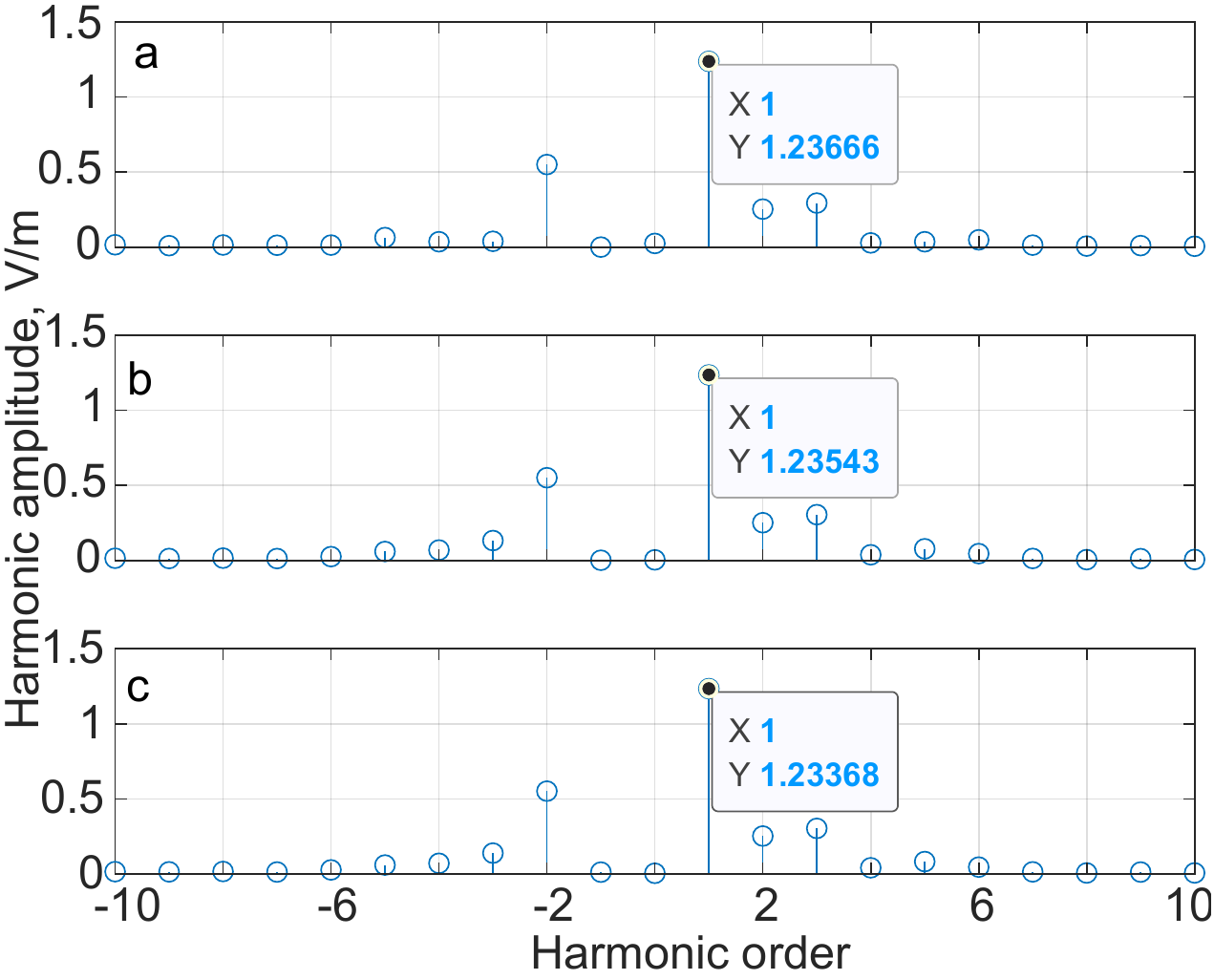}}
\caption{Amplitudes of scattered harmonics for designed metasurfaces with optimized  step-wise uniform sheets: (a) -- Design 1, (b) -- Design 2, and (c) -- Design 3.}
\label{fig:HarmonicsAmplitude}
\end{figure}

As a result of numerical optimization, we get a set of different solutions minimizing Eq.~\eqref{Eq:cost_function}. 
From them we select the most suitable one with reasonably decaying amplitudes of the evanescent surface wave harmonics, because high amplitudes of reactive fields near the array lead to smaller frequency bandwidth and higher losses. 
Figure~\ref{fig:HarmonicsAmplitude} shows an example of optimized mode amplitudes for the model of an infinite periodic structure. It is shown that the incident power is fully reflected to the anomalous direction ($n=1$) with nearly zero specular reflection and $n=-1$ order reflection.
The values of the discretized grid reactances are given in Table~\ref{tab:MainTable} as $Z_{1-8}$ for the corresponding design frequencies.

\begin{table}
\centering
\caption{Design parameters of the implemented anomalous reflectors. In case of the absence of an element in the sub-cell, the dog-bone element parameter is marked as "x".}
\begin{tabular}{||c | c | c | c||}
\hline
& 1 & 2 & 3 \\ [0.5ex] 
\hline\hline
$f_0$, GHz & 144.75 & 157.75 & 170.90 \\ 
\hline
$Z_1$, Ohm & $-319j$ & $-137j$ & $-1229j $\\
\hline
$Z_2$, Ohm & $-1686j$ & $-1010j$ & $43j$ \\
\hline
$Z_3$, Ohm & $-346j$ & $-876j$ & $-1074j$ \\
\hline
$Z_4$, Ohm & $-138j$ & $43j$ & $-926j$ \\
\hline
$Z_5$, Ohm & $-991j$ & $-833j$ & $-1250j$ \\
\hline
$Z_6$, Ohm & $-1721j$ & $-582j$ & $-2000j$ \\
\hline
$Z_7$, Ohm & $50j$ & $-775j$ & $-141j$ \\
\hline
$Z_8$, Ohm & $-1140j$ & $-1053j$ & $-2000j$ \\
\hline
$C_{x1}$, $\mu$m & 156 & 181.7 & 50 \\
\hline
$C_{x2}$, $\mu$m & 64 & 82 & 0 \\
\hline
$C_{x3}$, $\mu$m & 205 & 60 & 65 \\
\hline
$C_{x4}$, $\mu$m & 286.5 & 0 & 125 \\
\hline
$C_{x5}$, $\mu$m & 119.5 & 55 & 155 \\
\hline
$C_{x6}$, $\mu$m & 63.7 & 152 & x \\
\hline
$C_{x7}$, $\mu$m & 0 & 150 & 201.7 \\
\hline
$C_{x8}$, $\mu$m & 98 & 140 & x \\
\hline
$C_{y1}$, $\mu$m & 40 & 40 & 40 \\
\hline
$C_{y2}$, $\mu$m & 40 & 40 & 0 \\
\hline
$C_{y3}$, $\mu$m & 40 & 40 & 40 \\
\hline
$C_{y4}$, $\mu$m & 40 & 0 & 40 \\
\hline
$C_{y5}$, $\mu$m & 40 & 40 & 40 \\
\hline
$C_{y6}$, $\mu$m & 40 & 40 & x \\
\hline
$C_{y7}$, $\mu$m & 0 & 40 & 40 \\
\hline
$C_{y8}$, $\mu$m & 40 & 40 & x \\
\hline
$L_{x1}$, $\mu$m & 60 & 60 & 60 \\
\hline
$L_{x2}$, $\mu$m & 60 & 60 & 116.4 \\
\hline
$L_{x3}$, $\mu$m & 60 & 60 & 60 \\
\hline
$L_{x4}$, $\mu$m & 60 & 145 & 60 \\
\hline
$L_{x5}$, $\mu$m & 60 & 60 & 60 \\
\hline
$L_{x6}$, $\mu$m & 60 & 60 & x \\
\hline
$L_{x7}$, $\mu$m & 98 & 60 & 60 \\
\hline
$L_{x8}$, $\mu$m & 60 & 60 & x \\
\hline
$L_{y1}$, $\mu$m & 100 & 100 & 40 \\
\hline
$L_{y2}$, $\mu$m & 60 & 60 & $\lambda_0 /10$ \\
\hline
$L_{y3}$, $\mu$m & 100 & 60 & 40 \\
\hline
$L_{y4}$, $\mu$m & 100 & $\lambda_0 /10$ & 40 \\
\hline
$L_{y5}$, $\mu$m & 60 & 60 & 40 \\
\hline
$L_{y6}$, $\mu$m & 60 & 60 & x \\
\hline
$L_{y7}$, $\mu$m & $\lambda_0 /10$ & 60 & 80 \\
\hline
$L_{y8}$, $\mu$m & 60 & 60 & x \\
\hline
$\eta_{\rm eff}^{\rm sim, LL}$, $\%$ & 99.72 & 98.65 & 97.33 \\
\hline
$\eta_{\rm eff}^{\rm sim, Lossy}$, $\%$ & 90.78 & 91.81 & 90.76 \\
\hline
\end{tabular}
\label{tab:MainTable}
\end{table}

Finally, we make full-wave simulations with Ansys HFSS to validate the structure performance on the level of the  impedance sheet model, studying a super-cell with $Z_{\rm g}$ consisting of eight elements at the grounded dielectric layer. 
The structure is simulated as an infinite array along  $x$ and $y$ with the periodic boundary conditions under TE illumination realized with a Floquet port excitation. 
Knowledge of the magnitudes  of the Floquet harmonics for a given periodicity along $x$ found in simulations allows us to estimate the macroscopic reflection coefficient of the infinite structure \cite{macro1,macro2} and calculate the far-zone field reflected from a finite-size metasurface, whereas the reflection angle is defined in accordance with the model of diffraction on infinite periodic structures \eqref{Floquet_period}. 
Typically, the best numerically optimized solutions can realize the desired anomalous reflection with the efficiency at the level of 99.99\%, if material losses are neglected. 

\subsection{Implementation of elements in the super-cell}

Our next goal is to find proper geometries  of the metallic sub-cells that would realize the desired grid impedances $Z_k$. The grid impedance of a metallic patch or strip positioned on a grounded substrate can be determined using the circuit model \cite{Wang2018SystematicDesign}. To do that, we use the locally periodic approximation and simulate a single sub-cell using periodic boundary conditions in both $x$ and $y$ directions along the metasurface plane and find the reflection coefficient. The grid impedance of the meta-atom can be determined  from the reflection coefficients and the substrate parameters.
The incident angle in the simulation is defined as $\theta_{\rm i}$.
The input impedance of the structure can be expressed in terms of the simulated reflection coefficient $R$ as 
\begin{equation}
    Z_{\rm in} = \frac{1+R}{1-R}\eta_0,
\end{equation}
where $\eta_{\rm 0} = \sqrt{\mu_0 / \epsilon_0}$ is the free-space impedance, $\mu_0$ and $\epsilon_0$ are the
permeability and permittivity of free space, respectively  \cite{pozar2011microwave, Wang2018, Wang2018SystematicDesign}. 
$Z_{\rm in}$ is the input impedance of the parallel connection of the grid impedance $Z_{\rm g}$ and the grounded substrate impedance $Z_{\rm gd}$. Next, we extract the grid impedance using
\begin{equation}
    Z_{\rm g} = \frac{Z_{\rm in} Z_{\rm gd}}{Z_{\rm gd} - Z_{\rm in}},
    \label{eq:Z_g}
\end{equation}
where $Z_{\rm gd} = j \eta_{\rm d} \tan{k_{\rm d} d}$  and $k_{\rm d} = k_0 \sqrt{\epsilon_{\rm d}-\sin^2 \theta_{\rm i}}$ is the propagation constant in the dielectric substrate with $\eta_{\rm d}=\eta_0/\sqrt{\epsilon_{\rm d}-\sin^2\theta_{\rm i}}$ (for TE-polarized wave)  and the free-space wavenumber $k_0=\omega \sqrt{\epsilon_0\mu_0}$  \cite{pozar2011microwave}. 

\begin{figure}
\centerline{\includegraphics[width=1.0\columnwidth]{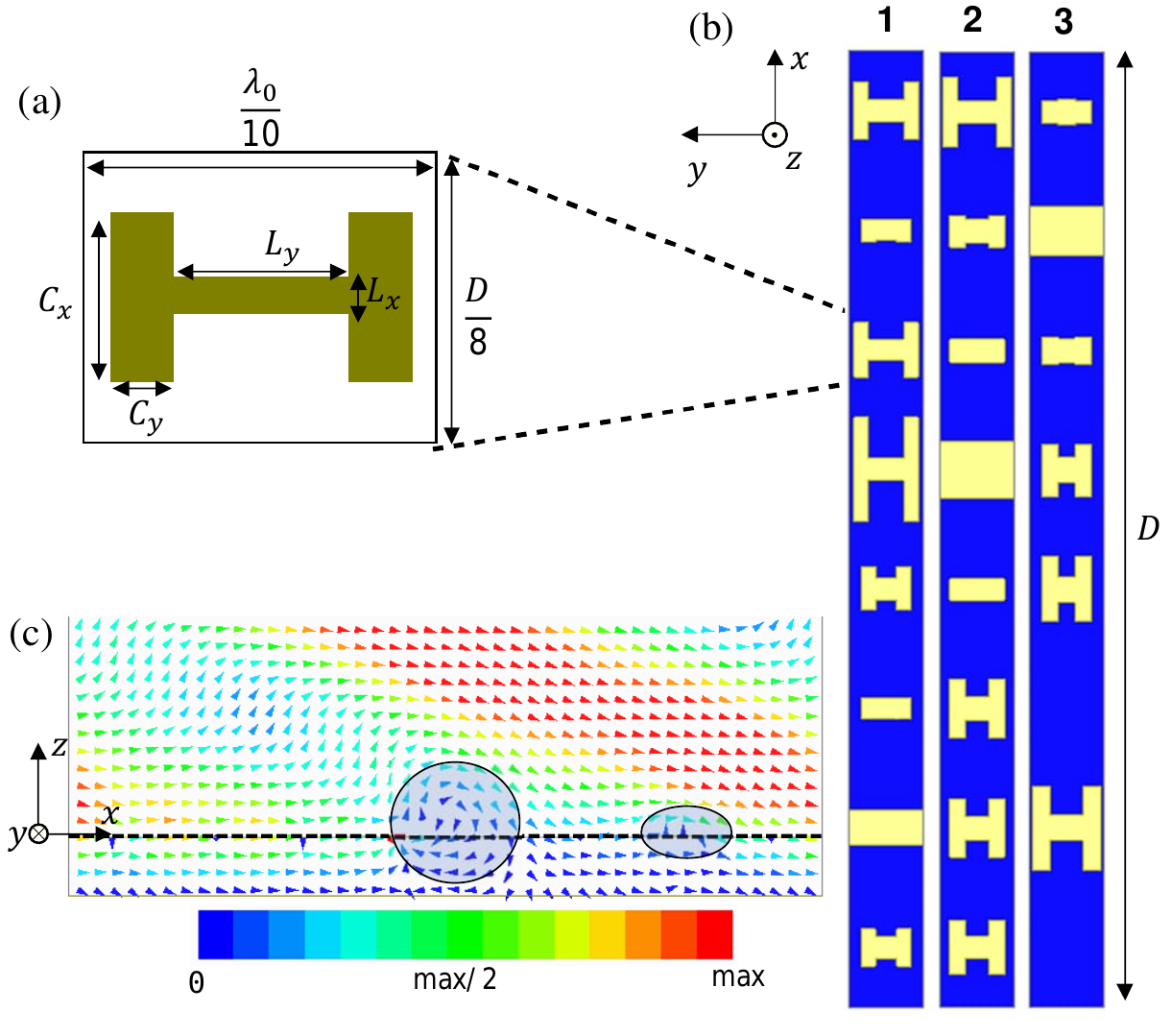}}
\caption{Geometry of the implemented metasurface: (a) -- a sub-cell with parameters varied in the impedance extraction; (b) -- final super-cells implemented for three targeted designs of perfect anomalous reflectors. (c) Poynting vector distribution at the $zx$ plane. The shaded circle regions show the regions of non-local power exchange between free space and the metasurface volume.  }
\label{fig:DogBoneTopology}
\end{figure}

\begin{figure}
\centerline{\includegraphics[width=0.9\columnwidth]{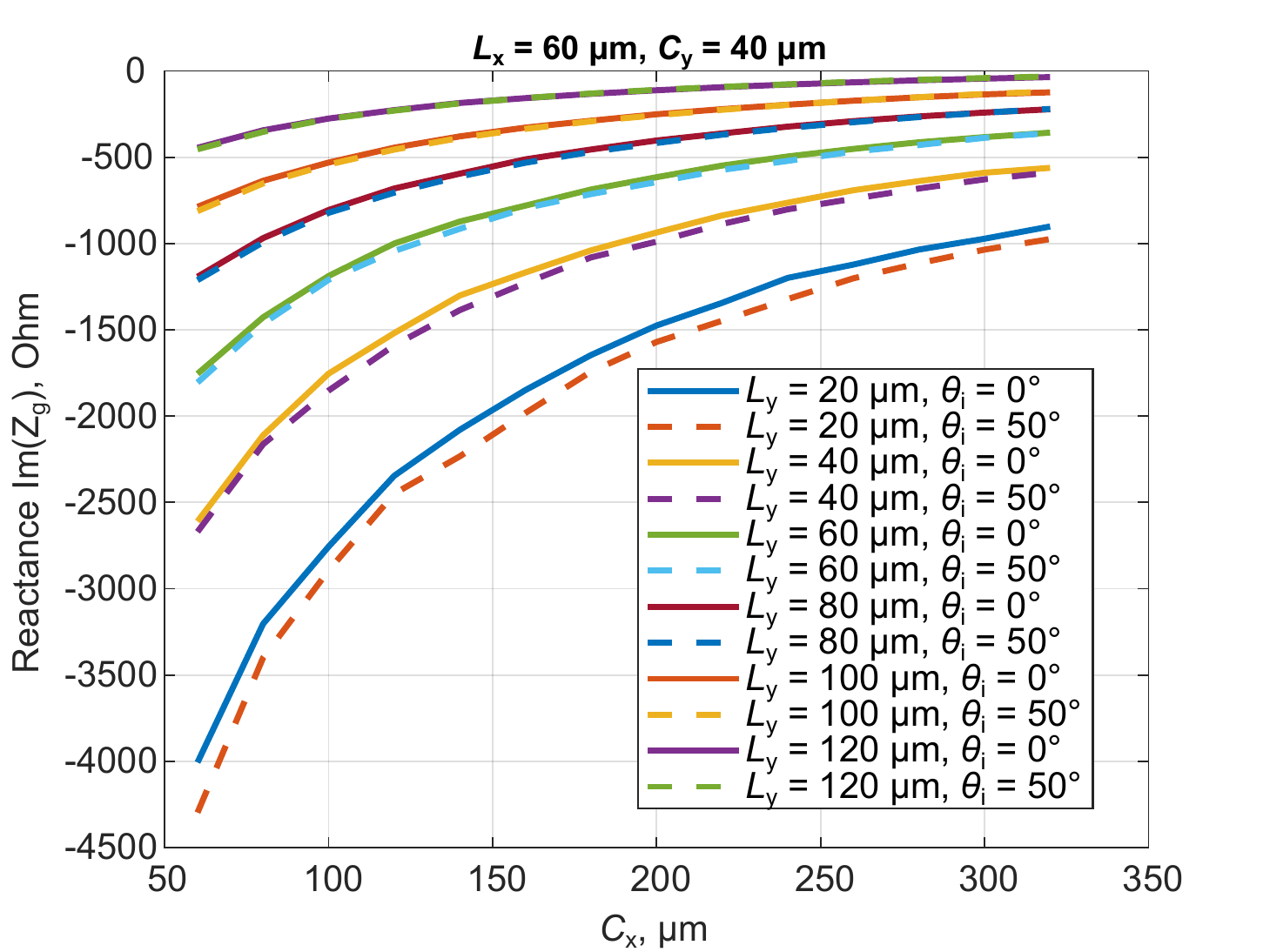}}
\caption{Extracted impedance values and the angular stability analysis.}
\label{fig:ExtraxtedZg}
\end{figure}

The simplest geometry of capacitive sub-cells is a rectangular patch, but such elements exhibit very different responses for oblique and normal incidences. 
This strong angular dependence leads to some impedance mismatch when the elements are implemented in the final super-cell, therefore requiring an additional global optimization of the whole super-cell. For this reason, we used the so-called {\it dog-bone} geometry for the sub-cell metal pattern, illustrated in Fig.~\ref{fig:DogBoneTopology}(a). This is one example of self-resonant grids \cite{Anderson} that are known to offer high angular stability of response from high-impedance surfaces, e.g.  \cite{Sim,Schu,Javad}.
Varying the element's dimensions we obtain  corresponding parameters of the sub-cell to realize  the required impedance values. 
Results for the extracted reactance $Z_{\rm g}$ for  different structural parameters for Design 1 are presented in Fig.~\ref{fig:ExtraxtedZg}. 
It is worth noting a very high level of angular stability comparing the results for the normal and oblique incidences. 
% The overall performance enhancement after the final tuning of the super-cell was not dramatic and granted just for 1-3\% improvement of the anomalous reflection efficiency.  
The obtained values of the implemented elements' parameters are presented in Table~\ref{tab:MainTable}, and the top view of the implemented super-cells is shown in Fig.~\ref{fig:DogBoneTopology}(b). 
Some of the elements require rather large negative values of reactance, and our analysis shows that these elements can be substituted with simply open sub-cells without any element presented. Some other elements require a small positive reactance which is implemented as inductive strips with the width $L_x$, that is, the length is equal to the overall $y$ periodicity of the metasurface $L_y = \lambda_0/10$.

The final structures are simulated as infinite periodic structures. The results of simulations in the absence and presence of material losses are given in Table~\ref{tab:MainTable} as $\eta_{\rm eff}^{\rm sim, LL}$ (with lossless metal) and $\eta_{\rm eff}^{\rm sim, Lossy}$ (with gold layers of 200~nm thickness), respectively. 
The results show that the reflection efficiency slightly ($\sim 1-3$~\%) degrades after the implementation of a realistic structure, in comparison with the performance of the optimized sheet-impedance model. 
When metal losses are taken into account, the efficiency degrades towards $\sim 90$ \%, mainly due to excited non-propagating surface waves along the metasurface. The efficiency can be improved if the metal deposition thickness increases (reducing the Ohmic loss in metal). 

Figure~\ref{fig:DogBoneTopology}(c) shows the simulated Poynting vector distribution of Design 1 in the $xoz$ plane. It is clear that in the regions marked by shaded circles, there is  non-local power exchange between the metasurface volume and free space. The power flows into the metasurface (virtual loss) and then  it is re-radiated into free space (virtual gain).  For the other two implemented designs we see a qualitatively similar  picture showing non-locality of power reflection. Note that although the given examples shows non-local reflection and only global power conservation, the developed design method is general and can find both local and non-local solutions.

\section{Experimental validation}

In order to fabricate the designed anomalous reflectors we used a standard cleanroom photolithography manufacturing process, with 4-inch quartz wafers for the substrate. 
% Due to the holders used in the electron-beam lithography machine, there is a gap between the evaporated metal area and the edges of the wafer, thus the absence of an Ohmic contact with the ground plane is guaranteed. However, holders decrease the printable area size by several millimeters from the wafer edge. 
We used AZ514E photoresist, 5~nm Ti and 200~nm Au for both the front side patterning and the ground plane. This Au layer thickness was selected in order to reasonably minimize the resistive losses, as they may significantly affect the device performance. 

\begin{figure*}
\centerline{\includegraphics[width=1\textwidth]{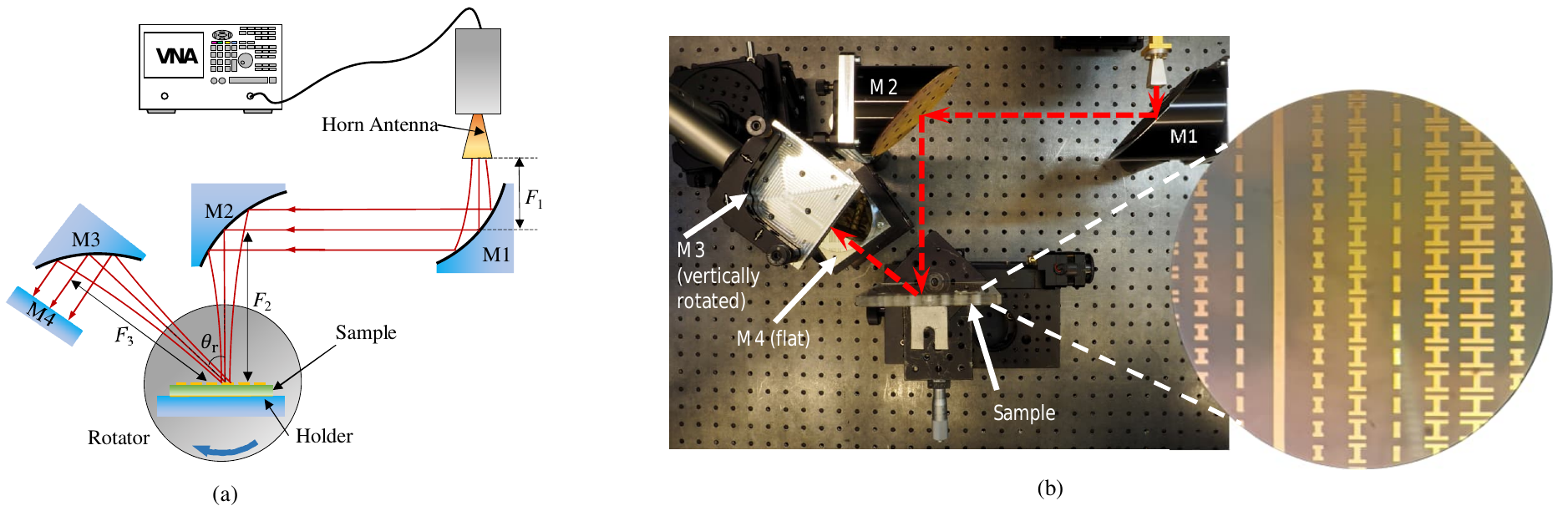}}
\caption{(a) Schematics and (b) photo of the experiment setup and structure details of Design 1 under microscope.}
\label{fig:ExperSetUp}
\end{figure*}

For experimental validation, we follow a similar procedure as presented in \cite{Wang2018}. A  quasi-optical set-up is used for collimation of the incident illumination from the horn antenna into a narrow spot on the metasurface sample under study. 
This allows us to effectively measure the macroscopic reflection coefficient as a parameter describing response from the corresponding infinite structure, avoiding the effects of scattering by the edges. 

The experimental set-up is shown in Fig.~\ref{fig:ExperSetUp}(a). 
The used vector network analyser (VNA) is Keysight PNA Network Analyzer N5225A (10 MHz -- 50 GHz) with WR 5.1 VDI extension units operating at 140 -- 220 GHz. A WR-5 rectangular horn antenna Elmika RHA-015E with 22 dBi directivity was used  as the source antenna. 
The set-up contains three $90^{\circ}$ parabolic mirrors M1, M2, and M3 with the corresponding focal distances $F_1 = 25.4$~mm, $F_2 = 152.4$~mm, and $F_3 = 127$~mm. Note that, the characterization of anomalous reflector usually needs two-port measurement system, i.e., two horn antennas are required in the normal and anomalous directions. Here, we use quasi-optical measurement system and time-gating method to simplify it as one-port system. The Gaussian beam emitted from the source Horn antenna is collimated by mirror M1 and converged by M2 towards the normal of sample. The beam is then reflected to the anomalous direction by the sample. Finally, the beam is reflected by flat mirror M4 and go back to the horn antenna along the same route. 
The efficiency of anomalous reflection is determined by measuring the reflection coefficient $S_{11}$. Note that in this method, the beam is reflected twice by the sample. Also, the spurious reflections from all the mirrors can be corrected by normalization. The normalization is made by measuring the reflection coefficient when the sample is replaced by a flat mirror rotated to $\theta_{\rm r}/2 = 25^{\circ}$. 
In order to estimate the retroreflection level for the normal incidence we measured the intensity of the reflected signal tuning the time-gating parameters for reflection from the sample. 
Another anomalously reflected harmonic ($n=-1$) was measured by upturning the sample. 
The last measurement is approximate, due to the fact that the measured area could not be exactly at the same position as the originally measured for the desired anomalously reflected mode. 
%, however not of the main focus of the current experimental demonstration.

\begin{figure*}
\centerline{\includegraphics[width=1\textwidth]{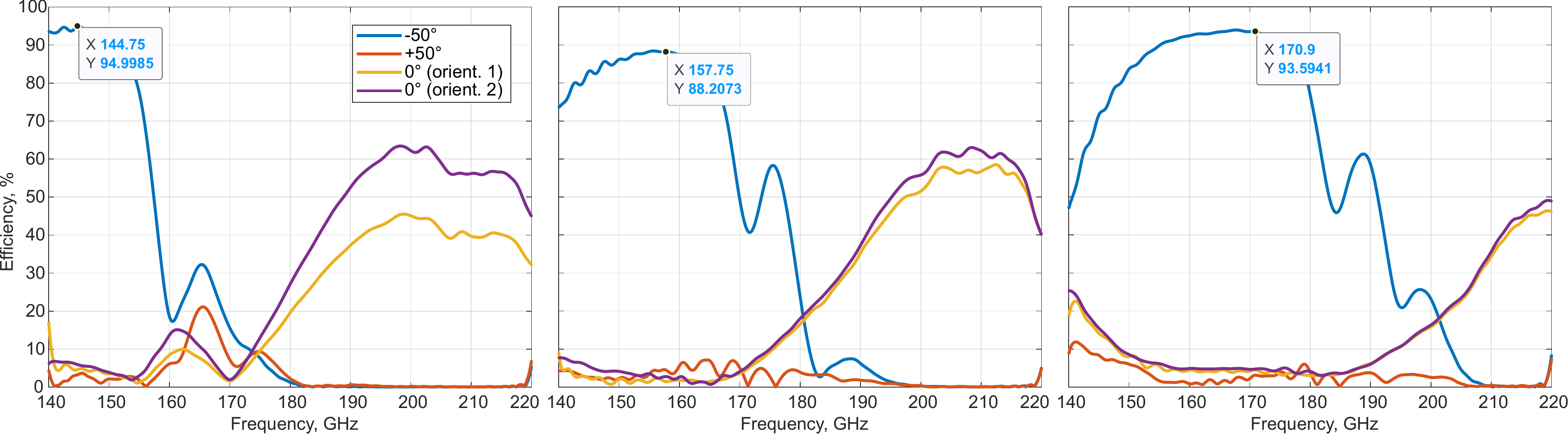}}
\caption{Reflection efficiency measured with the quasi-optical method. Curver marked ``$-50^{\circ}$'' corresponds to the desired anomalously reflected mode $n = -1$; Curve ``$+50^{\circ}$'' is for the same mode  measured after flipping the sample; ``$0^{\circ}$ (orient. 1)'' curve shows the retroreflection level for the normal incidence corresponding to the orientation of the sample applied while measuring the case with``$-50^{\circ}$'' and ``$0^{\circ}$ (orient. 1)'' corresponds to the same case, but for $+50^{\circ}$.}
\label{fig:Exper}
\end{figure*}

The experimental results are presented in Fig.~\ref{fig:Exper}. Here, the shown data point gives the efficiency measured at the targeted design frequency. These results  are in good agreement with the theoretically and numerically estimated values presented in Table~\ref{tab:MainTable}. 
For all three design cases, one can estimate the operational band of effective anomalous reflection. 
As a practically acceptable level, we define a region where the measured anomalous reflection efficiency is above 50\%. 
Due to the frequency limit of the used extender (140~GHz) we could not observe the whole operational bands for Designs 1 and 2, therefore the lower limit of the operational band in these cases is in fact wider than defined on the experimentally validated data. 
Design 1 provides more than half-power anomalous reflection efficiency in the range 140 -- 157.4~GHz; Design 2 grants it in the range 140 -- 170~GHz; and Design 3 covers the range 140.6 -- 183.1~GHz.

\section{Conclusions}

We have presented a method for the design and implementation of metasurfaces with advanced functionalities based on optimizations of discretized sheet  impedance profiles. 
The developed fast numerical optimization approach  allows us to directly find useful solutions without the  need of further discretization or detailed numerical tuning of the geometry of the implemented sub-cells. 
It is worth noting  that the proposed method is general, and not limited to the considered scenario of anomalous reflection. For example, it can be used to design perfect anomalous reflectors with larger reflection angles ($\theta_{\rm r}>50^\circ$). By introducing more impedance elements in one super-cell, it is also possible to realize dynamical beam steering with high efficiency. Such functionality cannot be realized by meta-gratings with sparse elements. Moreover, the method can be directly applied to the design of various beam-shaping surfaces,  splitters with phase control of all beams, and other metasurface devices, just by modifying the optimization objective function.  We consider this method very suitable for the development of optimized reconfigurable intelligent surfaces because by adding tunable components  to the sub-cells, the beam can be dynamically manipulated.

The considered example of  practical implementation was targeted for a $50^{\circ}$ anomalous reflector operating in D-band, which is one of the first realizations of anomalous reflectors in this frequency range. 
The performance of the designed structure has been  first validated with a full-wave simulation of an infinitely periodic reflector. 
Performance of the experimentally realized reflector has been confirmed by quasi-optical measurements that emulate the response of the infinite structure, allowing measurements of the macroscopic reflection coefficients for excited Floquet modes. The experimental results fully agree with the theoretical estimations and simulations.

\appendix
\section{Derivation details}
\begin{figure}
\centerline{\includegraphics[width=0.32\textwidth]{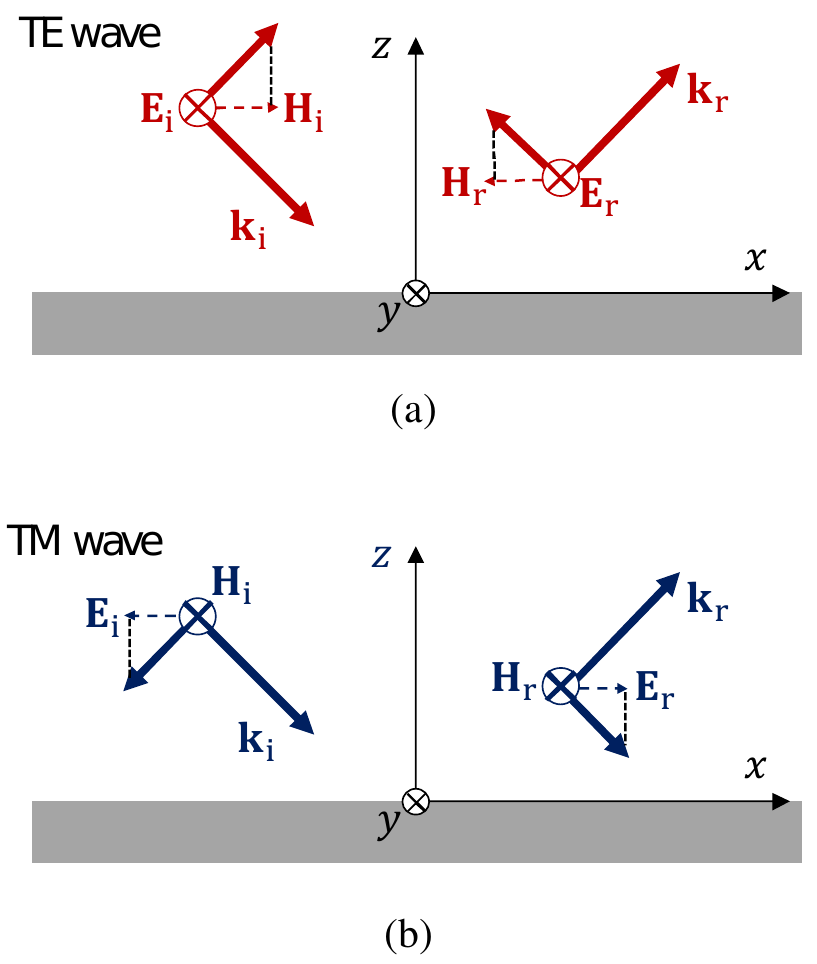}}
\caption{ Definitions of the coordinate system and field orientations for TE- and TM-polarized waves.}
\label{fig:Coordinate}
\end{figure}

Here, we show derivations of Eqs.~(\ref{eq: gamma1}) and (\ref{eq: gamma2}).  The total tangential electric and magnetic  fields are the sum of  tangential incident and reflected fields on the metasurface plane ($z=0$),
\begin{equation}
    \mathbf{E}_{\rm tot}=\mathbf{E}_{\rm i}+\mathbf{E}_{\rm r}, \label{eq: append1}
\end{equation}
and
\begin{equation}
    \mathbf{H}_{\rm tot}=\mathbf{H}_{\rm i}+\mathbf{H}_{\rm r}. \label{eq: append2}
\end{equation}
Assuming a TE-polarized wave incident on the metasurface, the tangential  magnetic field is related with electric field by the free-space wave admittance matrix:
\begin{equation}
    \hat z \times \mathbf{H}_{\rm i}=\mathbf{Y}_0\cdot\mathbf{E}_{\rm i}, \quad  \hat z \times \mathbf{H}_{\rm r}=-\mathbf{Y}_0\cdot\mathbf{E}_{\rm r}. \label{eq: append3}
\end{equation}
Note that the tangential magnetic field flips its sign after reflection, orienting along $-x$ direction, as shown in Fig.~\ref{fig:Coordinate}(a). 
The total tangential fields are related by the impedance boundary condition: 
\begin{equation}
    \mathbf{Y}_{\rm tot} \cdot \mathbf{E}_{\rm tot}= \hat z \times \mathbf{H}_{\rm tot}. \label{eq: append4}
\end{equation}
Substituting Eqs.~(\ref{eq: append1}), (\ref{eq: append2}), and (\ref{eq: append3}) into (\ref{eq: append4}), we can obtain the relation between the incident and reflected electric fields: 
\begin{equation}
    \mathbf{E}_{\rm r}=(\mathbf{Y}_{\rm tot}+\mathbf{Y}_{0})^{-1} \cdot (\mathbf{Y}_0-\mathbf{Y}_{\rm tot})\cdot  \mathbf{E}_{\rm i} =\mathbf{\Gamma}_{\rm TE} \cdot \mathbf{E}_{\rm i} .\label{eq: gamma3}
\end{equation}

For the TM polarization, we define the reflection matrix in terms of tangential magnetic fields $\mathbf{H}_{\rm r}=\mathbf{\Gamma}_{\rm TM}\cdot\mathbf{H}_{\rm i}$. In this case, it is more convenient to relate the tangential electric and magnetic fields via the wave impedance as
\begin{equation}
    \mathbf{E}_{\rm i}=\mathbf{Z}_0\cdot (\hat z\times \mathbf{H}_{\rm i}), \quad \mathbf{E}_{\rm r}=-\mathbf{Z}_0\cdot(\hat z\times \mathbf{H}_{\rm r}). \label{eq: append5}
\end{equation}
Substituting Eqs.~(\ref{eq: append1}), (\ref{eq: append2}), and (\ref{eq: append5}) into (\ref{eq: append4}), we obtain
\begin{equation}
    \mathbf{H}_{\rm r}=(\mathbf{Z}_{\rm tot}+\mathbf{Z}_{0})^{-1} \cdot (\mathbf{Z}_0-\mathbf{Z}_{\rm tot})\cdot  \mathbf{H}_{\rm i} =\mathbf{\Gamma}_{\rm TM} \cdot \mathbf{H}_{\rm i} .\label{eq: gamma5}
\end{equation}

\section*{Acknowledgment}

This work was supported in part by the European Commission through the H2020 ARIADNE project under grant 871464 and by the Academy of Finland under grant 345178. 

We acknowledge the use of facilities of Aalto University at OtaNano -- Micronova Nanofabrication Centre. The authors would like to thank 
Prof. Sami Franssila, 
Dr. Ville Jokinen, 
Dr. Victor Ovchinnikov, 
Dr. Mikhail Omelyanovich,
Dr. Andrey Generalov,
Dr. Irina Nefedova, and 
Mr. Pouyan Rezapoor
for their useful comments and discussions related to microfabrication and mm-wave measurement issues.

\bibliographystyle{IEEEtran}

\bibliography{references}

\end{document}